# Properties of Shape-Engineered *Phoxonic* Crystals: Brillouin-Mandelstam Spectroscopy and Ellipsometry Study

Chun Yu Tammy Huang[1], Fariborz Kargar[1,*], Topojit Debnath[2], Bishwajit Debnath[2], Michael D. Valentin[3], Ron Synowicki[4], Stefan Schoeche[4], Roger K. Lake[2], Alexander A. Balandin[1,*]

[1]Phonon Optimized Engineered Materials (POEM) Center, Department of Electrical and Computer Engineering and Materials Science and Engineering Program, University of California, Riverside, California 92521 USA

[2]Laboratory for Terascale and Terahertz Electronics (LATTE), Department of Electrical and Computer Engineering, University of California, Riverside, California 92521, USA

[3]Sensors and Electron Devices, CCDC Army Research Laboratory, Adelphi, Maryland 20783 USA

[4]J.A. Woollam Co., Inc., Lincoln, Nebraska 68508 USA

* Corresponding authors: (F.K) fkargar@ece.ucr.edu, (A.A.B) balandin@ece.ucr.edu ; web-site: http://balandingroup.ucr.edu/

**Abstract**

We report the results of Brillouin-Mandelstam spectroscopy and Mueller matrix spectroscopic ellipsometry of the nanoscale "pillar with the hat" periodic silicon structures, revealing intriguing phononic and photonic properties. It has been theoretically shown that periodic structures with properly tuned dimensions can act simultaneously as *phononic* and *photonic* crystals, strongly affecting the light-matter interactions. Acoustic phonon states can be tuned by external boundaries, either as a result of phonon confinement effects in individual nanostructures, or as a result of artificially induced external periodicity, as in the phononic crystals. The shape of the nanoscale pillar array was engineered to ensure the interplay of both effects. The Brillouin-Mandelstam spectroscopy data indicated strong flattening of the acoustic phonon dispersion in the frequency range from 2 GHz to 20 GHz and the phonon wave vector extending to the higher-order Brillouin zones. The specifics of the phonon dispersion dependence on the pillar arrays orientation suggest the presence of both periodic modulation and spatial localization effects for the acoustic phonons. The ellipsometry data reveal a distinct scatter pattern of four-fold symmetry due to nanoscale periodicity of the pillar arrays. Our results confirm the dual functionality of the nanostructured shape-engineered structure and indicate a possible new direction for fine-tuning the light-matter interaction in the next generation of photonic, optoelectronic, and phononic devices.

*Phoxonic* crystals (PxC) are referred to artificial materials, which exhibit simultaneous modulation of the elastic and electromagnetic properties within a single structure as a result of externally induced periodic boundaries.[1–4] One can consider PxC to be a structure with a concurrent dual functionality of the *phononic* crystals (PnC) and *photonic* crystals (PtC). Separately, PnCs[5–8] and PtCs[9–12] have been the subjects of intense theoretical and experimental investigations for the past two decades. Despite differences in intended applications, the structures share common physical characteristics. In PnC, a periodic modulation of the elastic constants and mass density define the phonon propagation while in PtC a periodic modulation of the dielectric constants define the photon propagation. The PnC and PtC arrays are tailored through construction of periodic lattices of holes or pillars, or a combination of both, with fine-tuned dimensions and unit cells. Since modulation of visible light and acoustic phonons of a certain energy range both require structure dimensions on the order of a few hundreds of nanometers one can envision a structure with the dual phonon – photon functionality. A proper selection of the periodic structure, with contrasting acoustic and optical properties, as well as a specific design of the lattice geometry and dimensions set up the common platform for PxCs, where novel phonon and photon characteristics and enhanced light-matter interactions are observed.[13–15] This approach has already been utilized in designing new type of optoelectronic devices such as phoxonic sensors[16–18] and optomechanical cavities.[13,19]

Engineering the phonon states of materials, *i.e.* changing the phonon properties by imposing spatial confinement in individual[20–23] or by inducing artificial periodicity as in PnCs,[5–8] has proven to be beneficial for thermal management, particularly at low temperatures,[24–27] acoustic filtering,[28] and wave guiding[29] applications. Acoustic phonons are the main heat carriers in electrically insulating and semiconducting materials, contribute strongly to the electron-phonon interactions in technologically important materials[30–34] and participate in the non-radiative generation-recombination processes.[35] The external periodicity of the pillar-based PnCs, additional to the crystal atomic periodicity, results in the zone-folding of acoustic phonons, and appearance of the new *quasi-optical* phonon polarization branches with non-zero energy at the Brillouin zone (BZ) center ($\omega(q = 0) \neq 0$ ).[36,37] The properties of quasi-optical phonons are substantially different from those of the fundamental longitudinal acoustic (LA) and transverse acoustic (TA) phonons, which have zero energy at the BZ center ($\omega(q = 0) = 0$) and a linear dispersion close to the BZ center. The quasi-optical phonons are similar to the true optical phonons but have the energies much lower than the optical phonons. The quasi-optical phonon modes are generally characterized by the hybridized vibrational displacement profiles. These modes change the phonon density of states (PDOS), and influence the thermal transport, especially at low temperatures, where the wave nature of phonons starts to dominate the phonon scattering processes. Generally, it is believed that the thermal conductivity in thin-film PnCs reduces due to the nanostructuring.[24,25,38] However, a recent theoretical study suggested that the thermal conductivity can be increased at low temperatures *via* fine-tuning of PnC dimensions and nanopatternings.[24,25]

Tuning the phonon dispersion in the PnC structures and in individual nanostructures can affect the optical and electrical properties of the material. Engineering the dispersion in PnCs in such a way that the phonon DOS attains it maximum or minimum within the energy required to trigger the carrier transition between the defect, *i.e.* trapping state and the conduction or valence band can result in either enhanced or suppressed G-R center recombination.[39] A recent experimental study found that modification of the phonon dispersion in the core-shell $GaAs_{0.7}Sb_{0.3}$/InP nano-pillar arrays affects the hot carrier relaxation in such structures.[34] In another example, opening up a band gap in a certain phonon frequency range with an accurate design of the hole PnCs can strongly influence the quasiparticle recombination lifetime in a superconductor.[40]

The periodic modulation of the dielectric constant in PtCs can lead to localization of electromagnetic waves, *e.g.* visible light, of a certain frequency range. In PtCs, a photonic energy stop band or a complete band gap can emerge, suppressing the electromagnetic wave propagation.[10] A common criterion for designing PtCs is utilization of the periodic structures with materials of the highest contrast in the refractive index.[10] Conventionally, PtCs, similar to PnCs, are fabricated in two configurations – the arrays of the air-holes[41] or the lattices of the pillars.[42] In the pillar-based structures, silicon nano-pillars are fabricated on a layer of a low refractive index material, *e.g.* $SiO_2$, to minimize the optical damping by the substrate.[43] However, recent studies demonstrated silicon-on-silicon PtCs with symmetric spheroidal-like nano-pillars, which are optically separated from the substrate and have low optical losses.[44] The coexistence of localized phonon and photon modes in these structures enhance the light-matter interaction, which can be beneficial for certain photonic, optoelectronic, or optomechanical devices.[13]

In this Letter, we report the results of Brillouin-Mandelstam spectroscopy (BMS), also referred to as Brillouin light scattering (BLS), and Mueller matrix spectroscopic ellipsometry (MM-SE) of the nanoscale "pillar with the hat" periodic silicon structures. The innovative idea of the study is to engineer the shape of the nanoscale pillar array in such a way that the phonon spectrum undergoes modification both due to periodicity of the arrays and phonon localization in the "hats" of the pillars. The larger diameters of the "hats" than that of the pillars are expected to result in better elastic decoupling from the substrates and resonant effects promoting acoustic phonon localization. From the other side, the periodicity of the structure was selected in the range ensuring the structure action as both PnC and PtC. Our results confirm the interplay of the localization and periodicity effects for phonons, as well the dual phonon and light functionality of the shape-engineered nano-pillar arrays. The described approach increases the range of tuning parameters available for optimization of PxC performance. The rest of the Letter is organized as follows. We describe the fabrication process of the shape-engineered silicon nanoscale pillar arrays, then present the results of BMS and MM-SE measurements, followed by our conclusions.

In Figure 1 (a-b) we present a schematic of the silicon "pillar with the hat" PxCs on a 500-µm thick silicon substrate with the associated dimensions and substrate crystallographic directions. The 500-nm tall pillars are arranged in a square lattice on a 1 mm × 1 mm (100) silicon chip along the <110> crystallographic directions with the inter-pillar center-to-center distance of $d = 500$ nm (see Supplementary Figure 1). The base of the silicon pillars is slightly larger than its end where it connects to the "hat" structure. The "hat" structure on top of the pillars has a symmetric shape with 443-nm lateral dimension (Figure 1-b). The unit cell of the PxC contains a quarter of each pillar with the space between them. Figure 1 (c-f) illustrates the step-by-step fabrication procedure of the PxCs. A thin layer of photoresist poly (methyl methacrylate) (PMMA 950 A2) was spin-deposited on a 500 µm thick (100) orientation silicon wafer and patterned in a square lattice arrays in the desired dimensions using the electron beam lithography (Figure 1-c). The silicon was then etched away using the silicon trench etch system (Oxford Cobra Plasmalab Model 100) at -120 °C under SF$_6$ and O$_2$ gas flow at 80 SCCM and 18 SCCM, respectively. Etching at low temperatures is required to slow down and control the process so that the structure will acquire the designer "pillar-with-the-hat" geometry. The obtained samples were characterized using the scanning electron microscopy (SEM) (Figure 2 (a-b)). It may appear from the side view of the SEM image shown in Figure 2 (a) that the adjacent pillars are connected to each other *via* their "hats". However, a magnified image of the pillars from the front view presented in Figure 2 (b) confirms that the nanostructures are completely separated as desired. The illusion of connected "hats" originates from the optical effects at a low magnification and a certain angle of view.

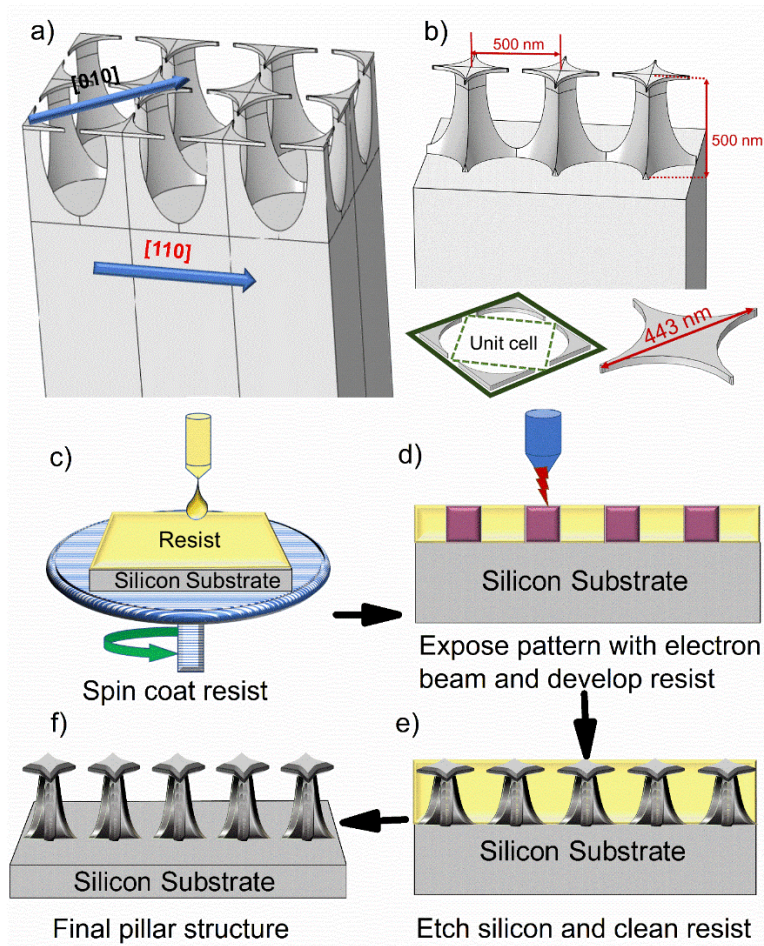

**Figure 1:** Schematic of the "pillars with the hat" structures on silicon substrate, showing (a) the crystallographic directions and the square lattice configuration, and (b) the geometry and features sizes of the nano-pillars. The periodic lattice consists of nano-pillars with 500-nm height and 500-nm pitch. The pillars have a symmetric "hat like" structure on top, which provides additional tunability for the phononic – photonic crystals. (c-f) Step-by-step illustration of the fabrication procedure.

We used BMS – an inelastic light scattering technique – in the backscattering geometry to measure the dispersion of the acoustic phonons in the energy range from 2 GHz to 20 GHz near the BZ center. BMS is a nondestructive optical tool which has been widely used to probe acoustic phonons and magnons in various types of material systems including transparent amorphous and opaque materials,[45–48] PnCs,[49,50] nanospheres,[51,52] membranes[53] and nanowires,[54] ferromagnetic[55,56] and antiferromagnetic materials.[57] The principles of BMS are similar to those of Raman spectroscopy. Raman spectroscopy is used to observe optical phonons with energies of one or two order of magnitudes higher than those of the acoustic phonons detectable with BMS. The details of the BMS measurement procedures are provided in the Methods, while the general approach is described in archival publications.[45,46,58,59] In BMS, two different mechanisms contribute to light scattering. Their relative strength depends on the optical transparency of the material system at the wavelength of the excitation laser.[45] If the material system is optically transparent, scattering by bulk of the material, involving "volumetric" phonons, is the dominant mechanism. If the material system is opaque or semi-transparent, the BMS spectrum is dominated by light scattering from the sample surface *via* the "ripple scattering" mechanism. In the latter case, only the in-plane component of the phonon momentum is conserved during the light scattering process. The magnitude of the in-plane component of the wave vector (or momentum) of the phonon participating in the scattering depends on the incident angle ($\theta$) of the laser light with respect to the normal to the sample's surface. It can be described as $q_{\parallel} = 4\pi \sin \theta / \lambda$.[45,46] Here, $\lambda = 532$ nm is the laser excitation wavelength. One can infer form this equation that by changing the incident light angle, the phonon wave vector can be selected and thus, the phonon dispersion can be obtained by carrying out experiments at various incident light angles. It should be noted that in PxCs, similar to opaque and semi-opaque materials, the light scattering is dominated by the surface ripple mechanism.[49]

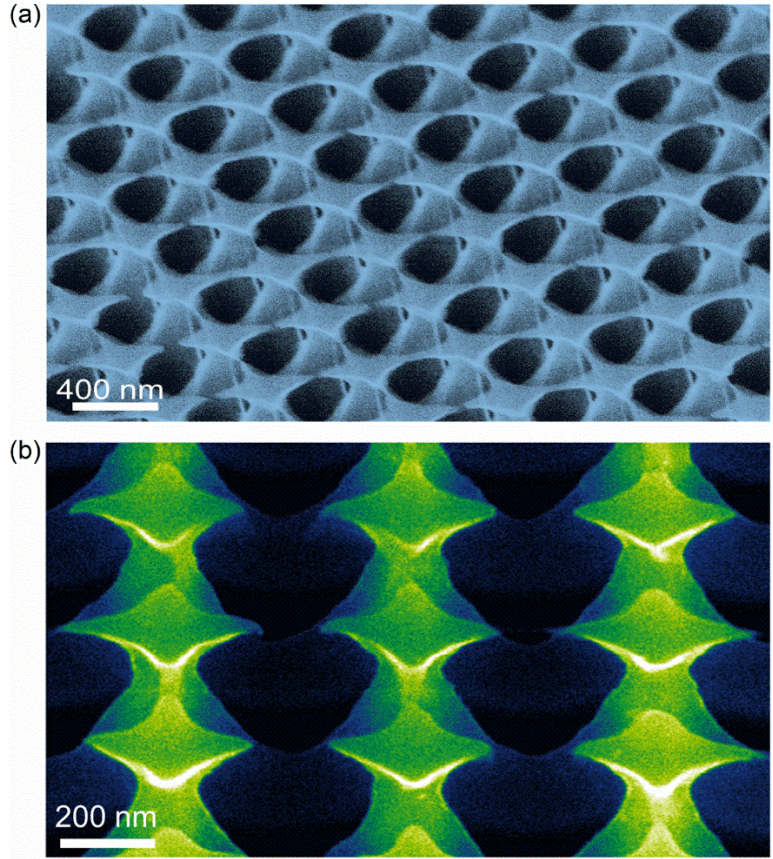

**Figure 2:** Scanning electron microscopy images of (a) the side view and (b) front view of the nano-pillar arrays. The pillars are separated from each other and not connected through their "hats".

Figure 3 (a) shows BMS data for the "pillar with the hats" at $\theta = 50°$ corresponding to the phonon wave vector of $q_{\parallel} = 18.1 \ \mu m^{-1}$ along the [110] crystallographic direction. All of the observed peaks are fitted using individual Lorentzian functions. The peaks at higher frequencies exhibits a rather large full width at half maximum (FWHM) when fitted with only one Lorentzian function. However, these broad peaks consist of two or more individual phonon peaks with frequencies too close to each other to resolve visually. In these cases, the peak deconvolution with several Lorentzian functions has been utilized. As one can see, there are nine peaks, including the peak shown in the inset, attributed to different phonon polarization branches. The peaks appear in the frequency range from 2 GHz to 20 GHz as a result of nanostructuring. The intensity of the peak at 2.2 GHz is higher than that of the rest of the peaks. For this reason, it has been plotted separately as an inset. It should be noted that the true bulk LA and TA phonons of silicon reveal two peaks at 90.6 GHz and 135.3 GHz as shown in Supplementary Figure 2.

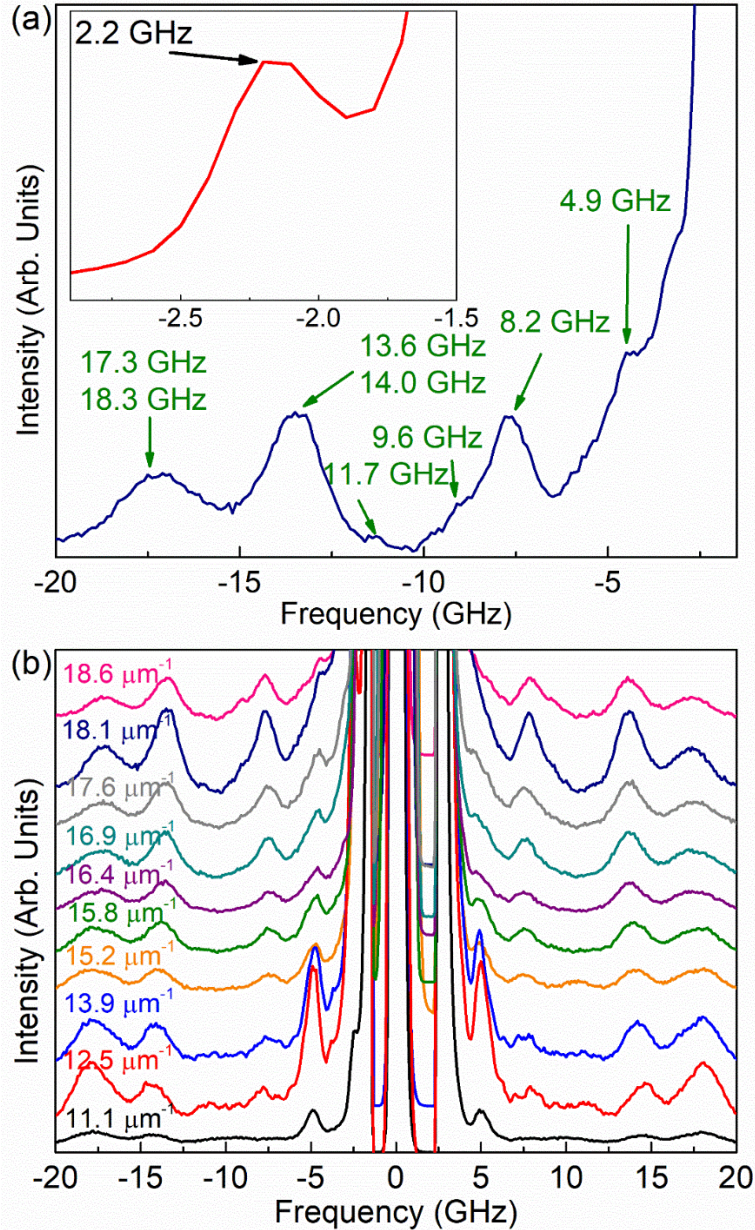

**Figure 3:** Brillouin-Mandelstam spectroscopy data for (a) the "pillar with the hat" structures accumulated at the incident light angle of 50°, corresponding to the in-plane phonon wave vector of $q_{\parallel} = 18.1\ \mu m^{-1}$. (b) The evolution of the phonon spectra with changing the probing phonon wave vector defined by the incident light angle.

In Figure 3 (b), we present the evolution of the phonon peaks for the "pillar with the hats" structure as a function of $q_{\parallel}$ varied by changing the incident light angle in the range of $28° < \theta < 52°$. This angle range corresponds to the phonon wave vectors range of $11.1\ \mu m^{-1} \le q_{\parallel} \le 18.6\ \mu m^{-1}$. All of the peaks observed in Figure 3 (a) are present in the spectra accumulated at different $q_{\parallel}$ with almost no observable changes in their spectral position, demonstrating that the frequency of the phonons do not depend on the phonon wave vector. This is an indication of the localized or standing phonon modes either within the "pillar with the hats" structure or in the substrate space between the pillars as a result of Bragg condition. The localized phonon modes possess zero group velocity ($v_g = \frac{\partial \omega}{\partial q} \sim 0$) and are hybrid in nature, revealing a complicated vibration displacement profile. The intensity of the peaks, however, changes for the different $q_\parallel$. For a semi-infinite opaque material, with the light scattering by surface ripple mechanism in the backscattering geometry, the total power, $dP$, scattered into the solid angle, $d\Omega$, for the light with polarization in the scattering plane, is described as $\frac{1}{P} \left( \frac{dP}{d\Omega} \right) = \frac{\omega_s^4}{\pi^2 c^4} A cos^4(\theta) R(\theta) \overline{e_z(0)^2}$.[45,60] Here $\omega_s$ is the frequency of the scattered light, $c$ is the speed of light in vacuum, $A$ is the area of the sample under illumination, $R(\theta)$ is the surface reflectivity which is by itself a function of the incident light angle $\theta$, and $\overline{e_z(0)^2}$ is the mean square displacement of the surface at sample's interface. As one can see, the intensity of the scattered light for a specific phonon polarization branch depends on the reflectivity at the incident angle of $\theta$, which now also depends on the orientation and geometry of the pillars.

The PxC sample in the current study is an array of pillars arranged in a square lattice and thus, possess a 4-fold rotational symmetry. This means that the results of BMS experiment would be the same as the one presented in Figure 3 (a) if the sample is rotated about the axis normal to the sample by $\alpha = 90$ degrees. The frequency of the phonon modes, which are localized within the "hat" or "pillar" structure should not depend on the crystallographic direction of the PxC. This fact provides a tool for distinguishing the phonon spectrum changes due to the localization from that due to periodicity of the structure. Practically, it can be done by conducting BMS experiments at orientation angles, $\alpha$, where $q_\parallel$ lies along directions other than the rotational symmetric directions. One should note that the orientation angle $\alpha$ is different from that of the incident laser light angle $\theta$. The former defines the direction and the latter determines the magnitude of the phonon wave vector in BMS experiment, respectively. Figure 4 presents the results of the BMS experiment at constant phonon wave vector $q_\parallel = 18.1 \ \mu m^{-1}$ along two different crystallographic directions of [110] ($\alpha = 0$, black curve) and [010] ($\alpha = 45°$, red curve). It is important to note that some phonon modes, which were observed along the [110] direction, have disappeared in the spectrum accumulated along the [010] direction. One can conclude that the modes, which are present in both spectra with the same frequency are spatially localized phonon modes. The modes that disappeared or changed their frequency can be the ones either resulting from the nano-pillar arrays periodicity, *via* elastic coupling through the substrate, or at least affected by the periodicity.

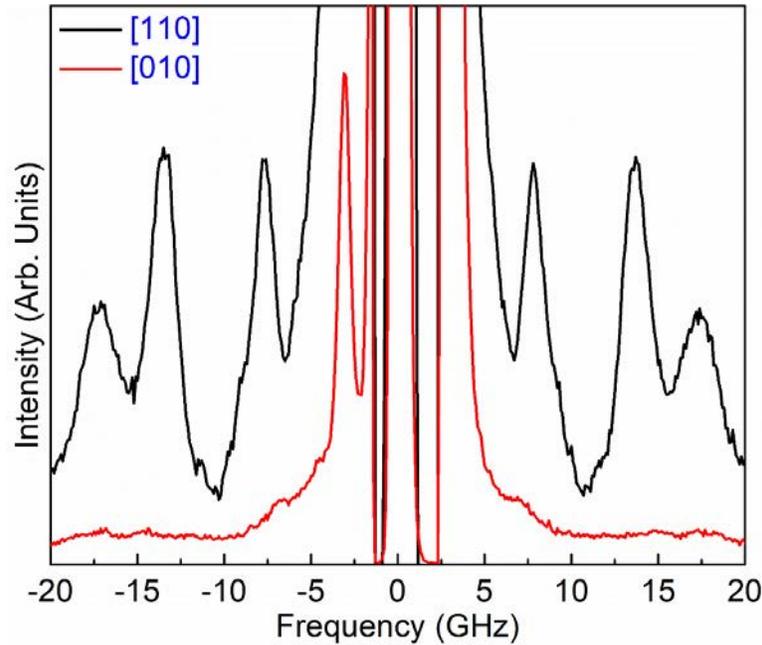

**Figure 4:** Brillouin-Mandelstam spectroscopy data for two different crystallographic directions in the same shape-engineered silicon nanoscale pillar array. Note that changing the orientation of the BMS scanning results in disappearance of some of the phonon peaks.

Imposing the artificial periodicity, *e.g.* adding holes or nano-pillars to the silicon substrate, changes the BZ geometry. In our case, since the "pillar with the hat" structures are arranged in a two-dimensional square lattice, the first BZ has a square geometry with the boundaries located at $\pi/a$, where $a = 500\ nm = 0.5\ \mu m^{-1}$ is the pitch, *i.e.* the distance between the central axis of two adjacent pillars along the [110] direction. Thus, the boundaries of the first BZ are located at $6.28\ \mu m^{-1}$ and $8.88\ \mu m^{-1}$, along the [110] and [010] directions, respectively. We conducted BMS experiments along the [110] and [010] crystallographic directions at various incident light angles ranging from $\theta = 18°$ to $\theta = 52°$, corresponding to $7.3\ \mu m^{-1} \leq q_\parallel \leq 18.6\ \mu m^{-1}$, respectively. The experimental data, accumulated in this range of $q_\parallel$ values, cover the most part of the 2$^{nd}$ and higher order BZ's along the [110] direction, and part of the 1$^{st}$ and higher order BZ's along the [010] direction. The spectral position of the observed BMS peaks along the [110] and [010] directions are plotted as a function of $q_{\parallel}$ in Figure 5 (a-b). The frequency of the peaks does not depend on the $q_{\parallel}$, exhibiting almost a flat dispersion throughout the 1st and higher order BZ's. The phonon modes, which have the same frequency in both plots, are marked with the same symbol and colors. These modes are likely confined within the "pillar with the hat" structures since their dispersion does not change as the periodicity changes by rotating the sample by $\alpha = 45°$.

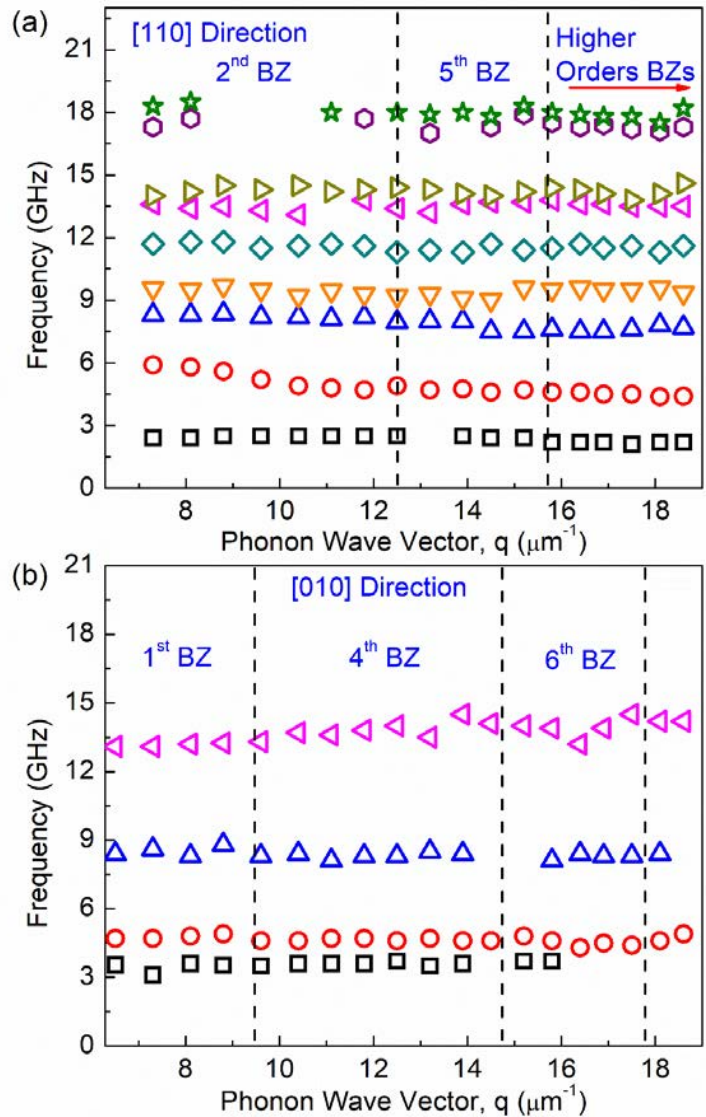

**Figure 5:** Phonon dispersion in the shape-engineered silicon nanoscale pillar array shown for (a) [110] crystallographic direction and (b) [010] crystallographic direction. Note that the number of observed phonon branches decreases from nine to four with the direction change.

To better understand the nature of the flat-band phonons, we calculated the phonon dispersion and displacement patterns for the silicon "pillar with the hat" structures using the finite element modeling (FEM) implemented in COMSOL Multiphysics package. The dimensions of the actual nanostructures have been determined from the SEM images under different angles. The schematic shown in Figure 1 (a-b) was used for the modeling. The details of the numerical simulations are described in the Methods section. The simulation results are presented in Figure 6 (a-b) for the [110] and [010] crystallographic directions. The calculated dispersion reveals numerous phonon polarization branches with the flat dispersion in both crystallographic directions, consistent with the experimental data. At higher frequencies, the phonon dispersion becomes a "spaghetti" of many phonon bands crossing and anti-crossing each other, making a direct comparison with the experiment impossible. It is important to note that these quasi-optical phonon modes are all hybridized in nature, which makes them BMS active, with the high degree of confidence. The hybrid modes have vibrational displacement components perpendicular to the probing phonon wave vector, which is a required condition for being observable by BMS.

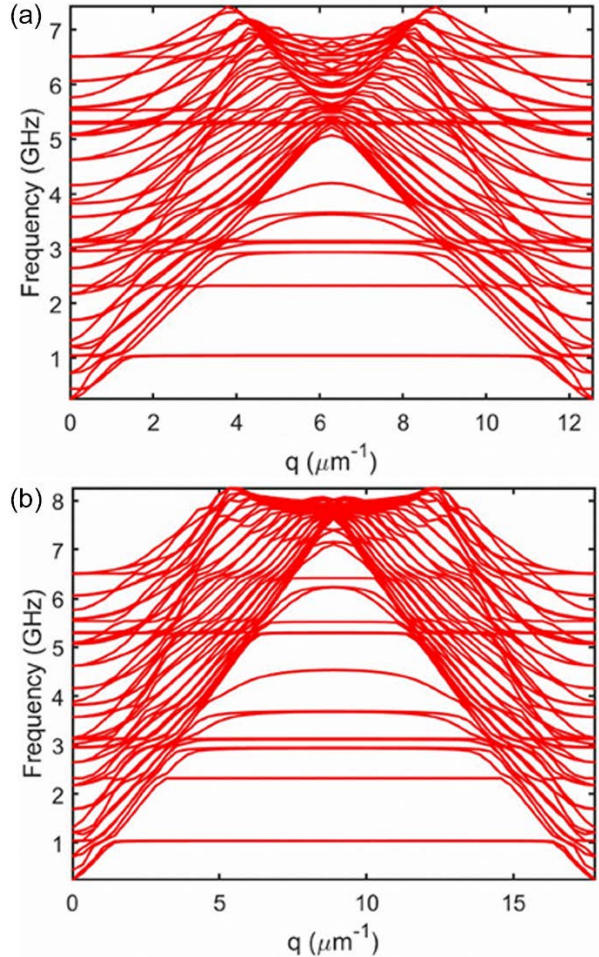

**Figure 6:** Calculated phonon dispersion for the pillar structure with similar geometry as the experimental samples. The data are shown for (a) [110] direction and (b) [010] directions.

The structures with similar periodicity and dielectric constant modulation are known to reveal photonic crystal properties.[1,11,42,44,61] To verify the modulation of the optical response we conducted ellipsometry measurements. Spectroscopic ellipsometry is a powerful and sensitive tool, which is now widely used for investigation of stacks of thin films, heterostructures, and nanostructured arrays with subwavelength gratings.[61] It is a technique, which is highly sensitive to surface plasmons and various optical effects revealed by metamaterials.[61,62] In this work, the optical characteristics of the shape-engineered silicon nanoscale pillar arrays were determined with the help of a dual-rotating compensator spectroscopic ellipsometer (RC2, J.A. Woollam Co.). Mueller matrix data was obtained for a full sample rotation and at multiple angles of incidence in specular reflection geometry. The details of the Mueller matrix ellipsometry experiment have been explained in detail elsewhere.[61]  Polar contour plots for a full azimuthal rotation of the sample and an angle of incidence of $\beta = 70°$ are presented in Figure 7. The radial component represents the photon energy between 0.73 eV and 5 eV while the angle represents the sample azimuth relative to the plane of incidence. Each Mueller matrix element of the "pillar with the hat" structures reveals a distinct scatter pattern of four-fold symmetry due to the symmetry and periodicity of the pillar array. The non-zero off-diagonal block Mueller matrix elements indicate strong cross-polarization for measurements that are not aligned with a symmetry axis of the array, and present a direct visualization of the PtC character of the sample. The observed features closely resemble the results reported for a photonic crystal formed by a square array of nanoparticles on glass and explained as Rayleigh - Woods anomalies.[61] The ellipsometry data provide additional confirmation that the designed shape-engineered silicon nanoscale pillar arrays can act simultaneously as phononic and photonic crystals.

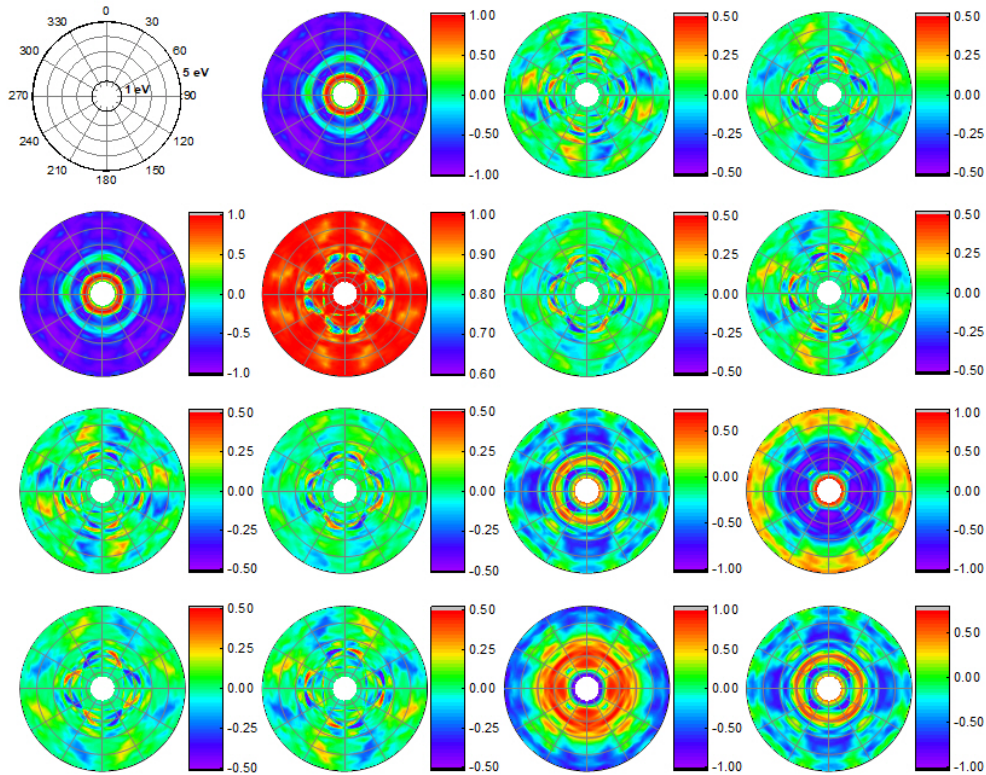

**Figure 7:** Polar contour plots of the normalized Mueller matrix spectroscopic ellipsometry data obtained at an angle of incidence of $\beta$=70° and for photon energies between 0.73eV and 5eV. The radial component represents the photon energy (in 1-eV steps) while the angle indicates the azimuthal orientation of the sample relative to the plane of incidence.

In conclusion, we used BMS and MM-SE to demonstrate that a specially designed nanoscale "pillar with the hat" periodic silicon structures, reveal the properties of both phononic and photonic crystals. Acoustic phonon states can be tuned by external boundaries, either as a result of phonon confinement effects in individual nanostructures, or as a result of artificially induced external periodicity. The shape of the nanoscale pillars was engineered to ensure the interplay of both effects. Our BMS data indicate strong flattening of the acoustic phonon dispersion in the frequency range from 2 GHz to 20 GHz and the phonon wave vector extending to the higher order BZs. The specifics of the phonon dispersion dependence on the pillar arrays orientation suggest the presence of both periodic modulation and spatial localization effects for the acoustic phonons. Our results suggest that smart engineering of the size and shape of the pillars enhances the tuning capability for phonon dispersion, which can be changed *via* spatial localization of confined phonons or pillar periodicity.

**METHODS**

**Brillouin-Mandelstam Spectroscopy:** BMS experiments were carried out in the backscattering geometry using a continuous wave (CW) solid state diode laser operating at the excitation wavelength $\lambda = 532$ nm. The incident light was $p$-polarized while there was no polarization selection for the scattered light. The laser light was focused on the sample using a lens with $f/\# = 1.4$. The angle of the incident light with respect to the normal to the sample has been changed from 16º to 52º using an automatic micro-rotational stage with an accuracy of 0.02º. This is required to select the phonon wave vector magnitude as described in the manuscript. The orientation of the sample has been adjusted using a small manual micro-rotational stage. The stage is required to select the crystallographic direction of the probing phonon wave vector. The scattered light from the sample was collected using the same lens and directed to the high resolution (3+3) pass Fabry-Perot interferometer (JRS Instruments).

**Mueller matrix spectroscopic ellipsometry:** The Mueller matrix represents the most general description of light interaction with a sample or optical element, including depolarization and scattering effects. It connects the incoming Stokes vector, $S_{in}$, with the outgoing Stokes vector, $S_{out}$

$$\begin{pmatrix} S_0 \\ S_1 \\ S_2 \\ S_3 \end{pmatrix}_{out} = m_{11} \begin{pmatrix} 1 & m_{12} & m_{13} & m_{14} \\ m_{21} & m_{22} & m_{23} & m_{24} \\ m_{31} & m_{32} & m_{33} & m_{34} \\ m_{41} & m_{42} & m_{43} & m_{44} \end{pmatrix} \begin{pmatrix} S_0 \\ S_1 \\ S_2 \\ S_3 \end{pmatrix}_{in} ,$$

with $S_0 = I_p + I_s$, $S_1 = I_p - I_s$, $S_2 = I_{+45°} - I_{-45°}$, and $S3 = I_{\sigma^+} + I_{\sigma^-}$ . where $I_j$ refers to intensity measurements for $p, s, +45°, -45°$, right-handed, and left-handed polarized light, respectively. Note the normalization of the Mueller matrix by the element $m_{11}$ which represents the total reflected intensity in the measurement. The off-diagonal block elements $m_{13}, m_{14}, m_{23}, m_{24}$ and $m_{31}, m_{41}, m_{32}, m_{42}$ are solely related to the Fresnel reflection coefficients $r_{sp}$ and $r_{ps}$. Therefore, non-zero values of these off-diagonal block elements directly indicate cross-polarization due to anisotropy or scattering in the sample. The RC2 spectroscopic ellipsometer is equipped with two continuously rotating compensators and determines the entire normalized Mueller matrix. No reference measurements need to be performed beyond a standard calibration routine. For the measurements performed here, the ellipsometer was equipped with focusing optics which reduce the measurement beam diameter from $\sim 3$ mm to $\sim 220$ µm. An automated rotation stage was carefully aligned to keep the measurement spot within the small structured sample area of $\sim$1x1 mm$^2$. Full Mueller matrix spectra for photon energies between 0.73 eV and 6.35 eV were obtained for multiple angles of incidence and a full sample rotation in azimuthal steps of 5°. An acquisition time of 10 s was used to obtain the full spectroscopic data set for each individual sample orientation.

**Finite-Element Method Simulations:** The pillars were modeled as a 2D periodic array, repeated in the x-y plane. The equation of motion for this phononic structure can be written from the second-order elastic continuum theory as $\rho(\partial^2 u(\mathbf{r})/\partial t^2) = \partial S(\mathbf{r})/\partial x_i$ where $\rho$ is the mass density and $u(\mathbf{r})$ is the displacement vector at coordinate $\mathbf{r}$. The stress tensor $S(\mathbf{r})$ can be obtained from displacement by $S_{ij} = C_{ijkl}\,\varepsilon_{kl}$, where $\varepsilon(\mathbf{r}) = [(\nabla\mathbf{u})^{\mathrm{T}} + (\nabla\mathbf{u})]/2$ is the elastic strain tensor. For this study, $C_{ijkl}$, the coefficients for silicon are assumed to be isotropic in both substrate and pillars. To obtain the solution of the elasticity equation in the frequency domain, the simulation geometry is discretized using a finite element scheme $\rho\omega^2 u = \nabla \cdot S$, where $\omega$ is the eigen frequency. A fixed boundary is applied at the bottom of the bulk substrate, and free surface boundary conditions are applied at all outer facets of the pillar as well as at the top surface, using $\varepsilon_{ij}n_j = 0$, where $n_j$ is the outward normal unit-vector. To simulate the effect of the additional hat-like structure on the phonon properties, we use a star-like protrusion on top of the pillar using the dimensions carefully extracted from the SEM images.

## Contributions

A.A.B. and F.K. conceived the idea of the study, coordinated the project, and contributed to the experimental and theoretical data analysis. C.-Y.T.H. fabricated the samples, conducted material characterization, BMS experiments, and contributed to the experimental data analysis. T.D. and B.D. performed computer simulations and contributed to analysis of the vibrational modes. M.D.V. assisted with the sample fabrication. R.S. and S.S. carried out the ellipsometry measurements and contributed to analysis of optical properties. R.K.L. supervised the phonon dispersion modeling and contributed to data analysis. F.K. and A.A.B. led the manuscript preparation. All authors contributed to writing and editing of the manuscript.

**Conflicts of Interest**

There are no conflicts to declare.

**Acknowledgements**

The work at UC Riverside was supported, by the DARPA project W911NF18-1-0041 on phonon engineered materials. The authors thank Dr. Hossein Taheri for the useful discussions, and acknowledge Dr. Ruben A. Salgado, Dr. Adane Geremew, Dr. Dong Yan, and Jacob S. Lewis for their help with material characterization at the UCR Nanofabrication Facility.